\documentclass[twocolumn,showpacs,amsmath,nofootinbib,prx,aps,amssymb,longbibliography]{revtex4-1}
\usepackage[T1]{fontenc}
\usepackage[utf8]{inputenc}
\usepackage{times}
\usepackage{color} 
\usepackage{array}
\usepackage{amssymb,amsmath}
\usepackage{amsbsy}
\usepackage[pdftex]{graphicx} 
\usepackage{bm}
\usepackage{float}
\usepackage{dcolumn}

\usepackage[unicode,breaklinks]{hyperref}
\hypersetup{
    unicode=true,
    plainpages=false, 
    colorlinks=true,
    linkcolor=blue,
    citecolor=blue,
    filecolor=black,
    urlcolor=blue
}
\usepackage{url}
\usepackage{verbatim}

\newcommand{\gdd}{g_\mathrm{dd}}
\newcommand{\add}{a_\mathrm{dd}}
\newcommand{\edd}{\epsilon_\mathrm{dd}}
\newcommand{\br}{\mathbf{r}}
\newcommand{\bx}{\mathbf{x}}

\newcommand{\Phidd}{\Phi_\mathrm{dd}}
\newcommand{\UD}{U_\mathrm{dd}}
\newcommand\gammaQF{\gamma_\mathrm{QF}}

\begin{document}
\title{Axial collective mode of a dipolar quantum droplet}
\author{P. Blair Blakie}
 \affiliation{Dodd-Walls Centre for Photonic and Quantum Technologies and Department of Physics, University of Otago, Dunedin 9016, New Zealand}
\date{\today}

\begin{abstract} 
In this work, we investigate the ground state properties and collective excitations of a dipolar Bose-Einstein condensate that self-binds into a quantum droplet, stabilized by quantum fluctuations. We demonstrate that a sum rule approach can accurately determine the frequency of the low energy axial excitation, using properties of the droplet obtained from the ground state solutions. This excitation corresponds to an oscillation in the length of the  filament-shaped droplet. Additionally, we evaluate the static polarizabilities, which quantify change in the droplet dimensions in response to a change in harmonic confinement.
\end{abstract}
 \maketitle
%%%%%%%%%%%%%%%%%%%%%%%%%%%%%%%%%%%%%%%%%%

%%%%%%%%%%%%%%%%%%%%%%%%%%%%%%%%%%%%%%%%%%
  
\section{Introduction}
A dipolar Bose-Einstein condensate (BEC) is made up of particles with a significant dipole moment, making dipole-dipole interactions (DDIs) important. Highly magnetic atoms have been used in experiments to create dipolar BECs \cite{Griesmaier2005a,Beaufils2008,Mingwu2011a,Aikawa2012a} and explore various phenomena arising from these interactions \cite{Lahaye_RepProgPhys_2009,Chomaz2023a}.
In the dipole-dominated regime, where DDIs dominate over short-ranged interactions, quantum droplets can be formed. These droplets are mechanically unstable at the mean-field level, but are stabilized against collapse by quantum fluctuations  \cite{LHY1957,Lima2011a,Lima2012a,Petrov2015a,Ferrier-Barbut2016a,Wachtler2016a,Bisset2016a}. Experiments using dysprosium \cite{Kadau2016a,Ferrier-Barbut2016a,Schmitt2016a} and erbium \cite{Chomaz2016a} atoms have prepared and measured various properties of these quantum droplets, which have a filament-like shape due to the anisotropy of the DDIs.

The extended Gross-Pitaevskii equation (EGPE) provides a theoretical description of the ground states and dynamics of quantum droplets, incorporating quantum fluctuation effects within a local density approximation. The collective excitations of  these quantum droplets are described by the Bogoliubov-de Gennes (BdG) equations, which can be obtained by linearizing the time-dependent EGPE about a ground state. Baillie \textit{et al.}~\cite{Baillie2017a} presented the results of numerical calculations of the collective excitations of a  dipolar quantum droplet (also see \cite{Pal2022a}). Additionally, a class of three shape excitations has been approximated using a Gaussian variational ansatz \cite{Wachtler2016b}. Experiments have measured the lowest energy shape excitation, corresponding to the lowest nontrivial axial mode, for a large trapped droplet \cite{Chomaz2016a}. Experiments with dysprosium droplets have also  measured the scissors-mode \cite{Ferrier-Barbut2018c}.   

In this paper,  we consider dipolar quantum droplet ground states and the $m=0$ (projection of angular momentum quantum number) collective excitations in free-space   and in trapped cases. We focus on a low energy axial collective excitation, which softens to zero energy to reveal the unbinding (evaporation) point for a free-space quantum droplet.
We use a sum rule approach to estimate the excitation frequency of this mode in terms of the ground state properties, including its static response to a small change in the axial trapping frequency. We compare the result of this sum rule method to the excitation frequency obtained by direct numerical solution of the BdG equations. We verify that the sum rule predictions are accurate over a wide parameter regime, including deeply bound droplet states, droplets close to the unbinding threshold, and in the trapped situation where the droplet crosses over to a trap-bound BEC.

 \section{Formalism}
\subsection{Ground states}
 
Here we consider a gas of magnetic bosonic atoms described by EGPE energy functional 
\begin{align}
E &= \int d\bx\, \psi^*\!\left[h_\mathrm{sp}+\tfrac12g_s|\psi|^2+\tfrac12\Phidd  +\tfrac25\gammaQF|\psi|^3\right]\psi,\label{Efunc}
\end{align} 
where 
\begin{align}
h_\mathrm{sp}&=-\frac{\hbar^2}{2M}\nabla^2+V(\mathbf{x}),\\
V(\mathbf{x})&=\frac{1}{2}M (\omega_x^2x^2+\omega_y^2y^2+\omega_z^2z^2),
\end{align} 
are the single particle Hamiltonian and the harmonic trapping potential, respectively.
Here $g_s = 4\pi \hbar^2 a_s/M$ is the coupling constant for the contact interactions, where $a_s$ is the $s$-wave scattering length.
The  potential
\begin{align}
\Phidd(\bx)=\int d\bx'\,\UD(\bx-\bx')|\psi(\bx')|^2,
\end{align}
describes the long-ranged DDIs, where the magnetic moments of the atoms are polarized along $z$ with
\begin{align}
\UD(\br) = \frac{3\gdd}{4\pi r^3}\left(1-\frac{3z^2}{r^2}\right).
\end{align}
Here $\gdd=4\pi\hbar^2\add/M$ is the DDI coupling constant, with $\add = M\mu_0\mu_m^2/12\pi\hbar^2$ being the dipolar length, and $\mu_m$ the atomic magnetic moment. The quantum fluctuations are described by the nonlinear term with coefficient $\gammaQF = \frac{32}3 g_s\sqrt{a_s^3/\pi}\mathcal{Q}_5(\edd)$ where 
\begin{align}
\mathcal{Q}_5(x)=\Re\left\{\int_0^1 du\,[1+x(3u^2 - 1)]^{5/2}\right\},
\end{align} \cite{Lima2011a} and $\edd \equiv \add/a_{s}$.   We constrain solutions to have a fixed number of particles $N$ and ground state solutions satisfy the EGPE 
\begin{align}
\mu\psi&= \mathcal{L}_\text{EGP}\psi,
\end{align}
 where $\mu$ is the chemical potential and we have introduced
 \begin{align}
 \mathcal{L}_\text{EGP}\psi&=\left(h_\mathrm{sp}+ g_s|\psi|^2+\Phidd  + \gammaQF|\psi|^3 \right)\psi.
 \end{align}
 For the results we present here we leverage the  algorithm developed in Ref.~\cite{Lee2021a} to calculate the ground states. 
 
\subsection{Excitations}
\subsubsection{Bogoliubov-de Gennes theory}
 The collective excitations are described within the framework of Bogoliubov theory.  These excitations can be obtained by linearising the time-dependent EGPE  $i\hbar\dot{\Psi}= \mathcal{L}_\text{EGP}\Psi$ around a ground state using an expansion of the form 
\begin{align}
\Psi \!=\!e^{-i\mu t}&\left[\psi(\mathbf{x}) \!+\!\sum_{\nu}\left\{ \lambda_\nu u_{\nu}(\mathbf{x})e^{-i\omega_{\nu}t}  
 \!-\! \lambda_\nu ^*v^*_{\nu}(\mathbf{x})e^{i\omega_{\nu}t}\right\} \right],
\end{align}
where  $\lambda_\nu$ are the (small) expansion coefficients.
The excitation modes  ${u_\nu,v_\nu}$ and frequencies $\omega_\nu$ satisfy the BdG equations
\begin{align}
 \begin{pmatrix}  \mathcal{L}_\text{EGP} - \mu + X & -X \\
      X &\! -(\mathcal{L}_\text{EGP} - \mu + X)\end{pmatrix}\!\begin{pmatrix} u_\nu \\ v_\nu\end{pmatrix}
     &= \hbar\omega_\nu \!\begin{pmatrix} u_\nu \\ v_\nu\end{pmatrix},\label{BdGX}
\end{align}
where $X$ is the exchange operator given by
\begin{align}
  Xf \equiv &g_s|\psi_0|^2f+\psi_0\int d\mathbf{x}'\,U_{\mathrm{dd}}(\mathbf{x}-\mathbf{x}') f(\mathbf{x}')\psi_0^*(\mathbf{x}')
   \\
  &+\tfrac32 \gamma_\text{QF} |\psi_0|^3f.\nonumber
\end{align}
More details of the Bogoliubov analysis of the EGPE can be found in Ref.~\cite{Baillie2017a}.  Here we will consider problems with rotational symmetry about the $z$ axis, which restricts the trap to  cases with $\omega_x=\omega_y$. This symmetry means that the excitations can be chosen to have a well-defined $z$-component of angular momentum, $\hbar m$, where we have introduced   $m=0,\pm1,\pm2,\ldots$ as the quantum number.

\subsubsection{Sum rule approach for lowest compressional mode}
We now wish to discuss a method for extracting the low energy axial compressional mode. We consider the mode excited by  the axial operator $\sigma_z=\int d\mathbf{x}\,\hat{\psi}^\dagger z^2\hat{\psi}$,  where $\hat{\psi}$ is the bosonic quantum field operator.
 The energy $(\hbar\omega_\mathrm{ub})$ of the lowest mode excited by $\sigma_z$  has an upper bound provided by the ratio $\sqrt{m_1/m_{-1}}$, where  we have introduced $m_p=\int d\omega\,(\hbar\omega)^pS_z(\omega)$ as the $p$-moment of  $S_z(\omega)$ \cite{BECbook}. 
Here 
\begin{align}
S_z(\omega)=\hbar\sum_{p}\left|\langle p |\sigma_z|0\rangle\right|^2\delta(E_p-E_0-\hbar\omega),
\end{align}
 is the zero temperature dynamic structure factor, with $|p\rangle$ being the eigenstate of the second quantized Hamiltonian  $H$ of energy $E_p$, and $|0\rangle$ denotes the ground state. %$E_\sigma|\sigma\rangle=H|\sigma\rangle$ 
 The $m_1$ moment can be evaluated using a commutation relation as
 \begin{align}
 m_1=\frac{1}{2}\langle 0|[\sigma_z,[H,\sigma_z]]|0\rangle=\frac{2N\hbar^2\langle z^2\rangle}{M}.
 \end{align}
The $m_{-1}$ moment can be evaluated as $m_{-1}=\frac{1}{2}\chi_{zz}$ (see \cite{BECbook}), where $\chi_{zz}$ is the static polarizability quantifying the linear response of $\sigma_z$  to the application of  the  perturbation $-\lambda \sigma_z$ to the Hamiltonian, i.e.~$\delta\langle \sigma_z\rangle=\lambda\chi_{zz}$.  Setting $\lambda\equiv -\frac{1}{2}M\,\delta\omega_{z}^{2}$, we see that this  perturbation is equivalent to a change in the axial trapping frequency, and hence the static polarizability can be evaluated as
\begin{align}
\chi_{zz}=-\frac{2N\delta\langle z^{2}\rangle}{M\delta\omega_{z}^{2}},
\end{align}
also see \cite{Menotti2002a,BECbook,Tanzi2019b}. Thus we obtain the following as an upper-bound estimate of the lowest mode frequency
\begin{align}
\omega_{\mathrm{ub}}=\sqrt{\frac{4N\langle z^2\rangle}{M\chi_{zz}}}.\label{omegaub}
\end{align}
A similar approach was used in Ref.~\cite{Ferrier-Barbut2018c} to estimate the scissors mode frequency of a dipolar droplet, but in that work the $z$-component of angular momentum was the relevant excitation operator.
 
 It is also of interest to understand the static polarizability quantifying the change in the transverse width (with operator  $\sigma_{\rho}=\int d\mathbf{x}\,\hat{\psi}^\dagger \rho^2\hat{\psi}$, where $\rho=\sqrt{x^2+y^2}$) to the axial perturbation, i.e.~
 \begin{align}
\chi_{\rho z}=-\frac{2N\delta\langle \rho^{2}\rangle}{M\delta\omega_{z}^{2}}.
\end{align}
Later we will consider both the \textit{axial} $\chi_{zz}$ and \textit{transverse} $\chi_{\rho z}$ static polarizabilities to characterize the quantum droplet behavior.

 \begin{figure*}[htbp!] 
   \centering
 \includegraphics[width=7in]{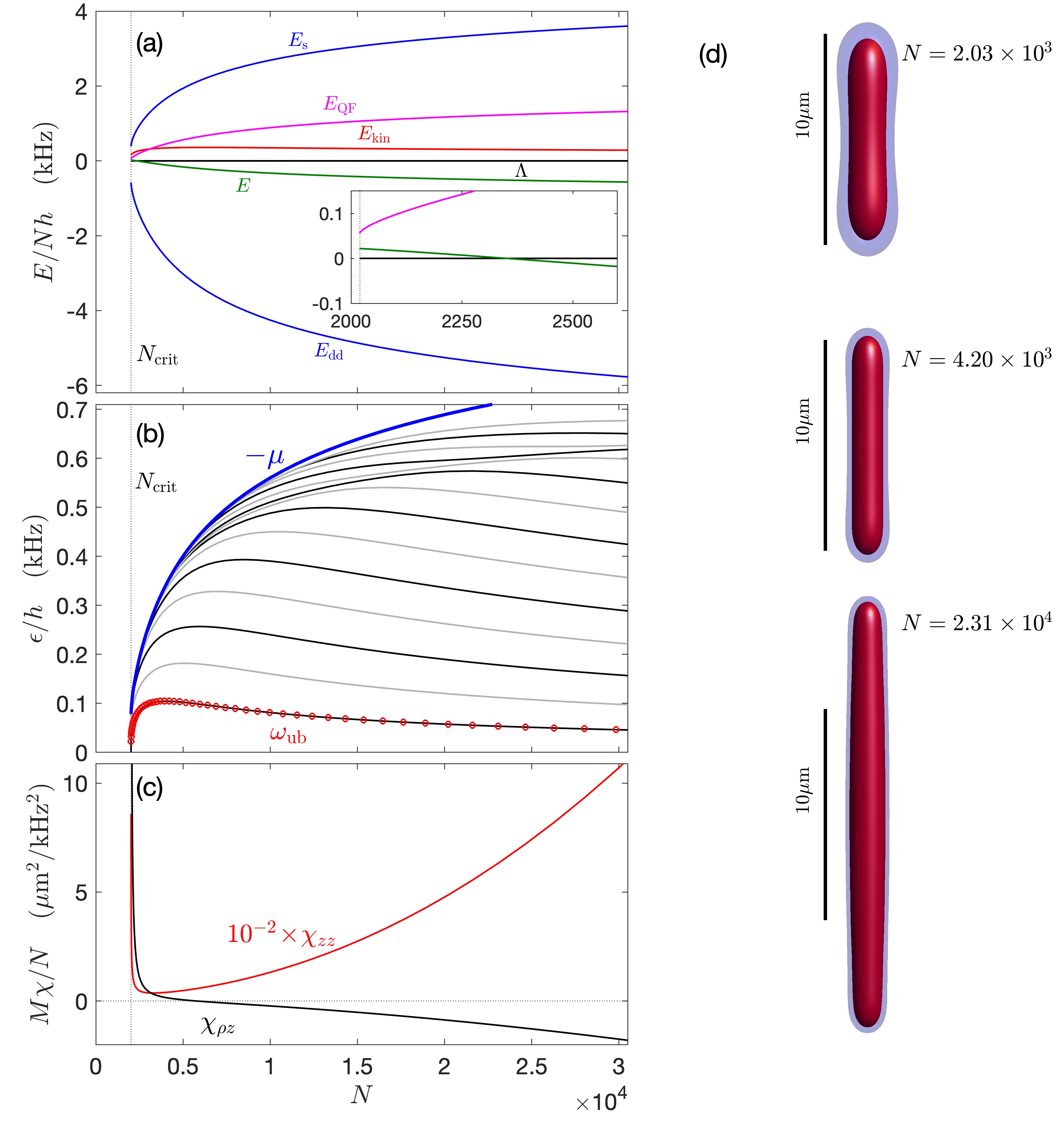}
   \caption{Free-space droplet ($V=0$) as a function of atom number $N$. (a) Components of energy and the virial $\Lambda$.  Inset shows a zoom in close to $N_\mathrm{crit}\approx2020$ to reveal where the total energy is positive. (b) Results from numerical solution of the BdG equations for the $m=0$ excitations that are even (black lines) and odd (grey lines) with respect to a reflection along $z$ through the droplet center. The approximate result $ \omega_\text{ub}$ obtained from (\ref{omegaub}) (red circles). The droplet binding energy, characterized by $-\mu$, is shown for reference (blue line).
   (c) The static polarizabilities $\chi_{zz}$ and $\chi_{\rho z}$.
Results for  $^{164}$Dy using $a_{\mathrm{dd}}=130.8\,a_0$ and $a_s=80\,a_0$.
(d) Density isosurfaces of droplet states for various $N$ values as labelled. Isosurfaces shown for  $10^{18}\,$m$^{-3}$ (blue) and $10^{19}\,$m$^{-3}$ (red).}
   \label{fig:fsdroplet}
\end{figure*}

 \section{Results for free-space droplets}
 \subsection{Energetics}
 In Fig.~\ref{fig:fsdroplet}  we consider a free-space (self-bound) $^{164}$Dy droplet for $a_s=80\,a_0$ (i.e.~$\epsilon_\mathrm{dd}=1.64$) as a function of $N$. For $N>N_\mathrm{crit}\approx2020$ a self-bound droplet solution exists\footnote{This value of $N_{\mathrm{crit}}$ is slightly higher than the value obtained  in Ref.~\cite{Baillie2017a} for the same physical parameters. This is because we use the result $mathcal{Q}_5(\epsilon_{\mathrm{dd}})$, rather than the approximation $1+\frac{3}{2}\epsilon_{\mathrm{dd}}^2$, to define $\gamma_{\mathrm{QF}}$.},  while for $N<N_\mathrm{crit}$ there is no droplet solution, and the atoms expand to fill all space.
  In Fig.~\ref{fig:fsdroplet}(a) we examine the energy of the free-space droplet, seeing that as $N$ approaches $N_\mathrm{crit}$ from above the total energy of the droplet becomes positive (for $N\lesssim2,350$, see inset). When the energy is positive the droplet is metastable.  We show some examples of the droplet states in  Fig.~\ref{fig:fsdroplet}(c) to give some context on the size and shape of the droplet as $N$ varies.
  
  We also consider the components of energy, defined by
  \begin{align}
 E_\text{kin} &=-\frac{\hbar^2}{2M}\int d\mathbf{x}\,\psi^*\nabla^2\psi,\\
 E_{\mathrm{pot}} &=\int d\mathbf{x}\, V(\mathbf{x})|\psi|^2,\\
 E_s &=\frac{1}{2}g_s\int d\mathbf{x}\,|\psi|^4,\\
 E_{\mathrm{dd}}&=\frac{1}{2}\int d\mathbf{x}\,\Phi_{\mathrm{dd}}(\mathbf{x})|\psi|^2,\\
  E_{\mathrm{QF}}&=\frac{2}{5}\gammaQF\int d\mathbf{x}\,|\psi|^5
 \end{align}
 as the kinetic, potential, contact interaction, DDI, and quantum fluctuation energy, respectively. 
For this free-space case $ E_{\mathrm{pot}}=0$. We observe that the DDI and contact interaction energies are the largest energies by magnitude, but have opposite sign. The quantum fluctuation term is necessary to stabilise the droplet solution against mechanical collapse driven by the interaction terms, but is typically significantly smaller in magnitude than either of the interaction energies. The kinetic energy is smaller than all other components except  when $N\to N_\mathrm{crit}$,  where it can exceed   $E_{\mathrm{QF}}$.
 
A virial relation for the system can be obtained by considering how the energy functional transforms under a scaling of coordinates (e.g.~see \cite{Dalfovo1999}). For the dipolar EGPE in a harmonic trap this relation is 
 \begin{align}
 \Lambda\equiv E_\text{kin}-E_{\mathrm{pot}}+\frac{3}{2}(E_s+E_\mathrm{dd})+\frac{9}{4}E_\text{QF}=0,
 \end{align}
 and provides a connection between the energy components   \cite{Triay2019a,Lee2021a}. In Fig.~\ref{fig:fsdroplet}(a) we show $\Lambda$ for reference.  This turns out to be a sensitive test of the accuracy of dipolar quantum droplet solutions \cite{Lee2021a}, and the results we present here typically have $|\Lambda/Nh|\ll10^{-3}\,$Hz.

 \subsection{Excitations}

  In Fig.~\ref{fig:fsdroplet}(b) we show the spectrum of $m=0$ modes in the droplet as a function of $N$. Excitations for other $m$ values have been examined in Ref.~\cite{Baillie2017a}. The lowest energy excitation branch for the quantum droplets is $m=0$ (this is not the case for vortex states, e.g.~see \cite{Lee2018a}).  We note that the excitations are measured relative to the condensate chemical potential. Thus excitations with $\hbar\omega_\nu>-\mu$ are unbounded by the droplet potential and are part of the continuum. For this reason we only calculate excitations with $\hbar\omega_\nu<-\mu$. From these results we observe that the lowest  $m=0$ mode goes soft (i.e.~approaches zero energy) as $N\to N_\mathrm{crit}$. This indicates the onset of a dynamical instability of the self-bound state. We also show the excitation frequency bound $\omega_\mathrm{ub}$ obtained from ground state calculations according to Eq.~(\ref{omegaub}). This is seen to be in good agreement with the BdG result for the lowest excitation over the full range of atom numbers considered.

 \begin{figure*}[htbp!] 
   \centering
 \includegraphics[width=7in]{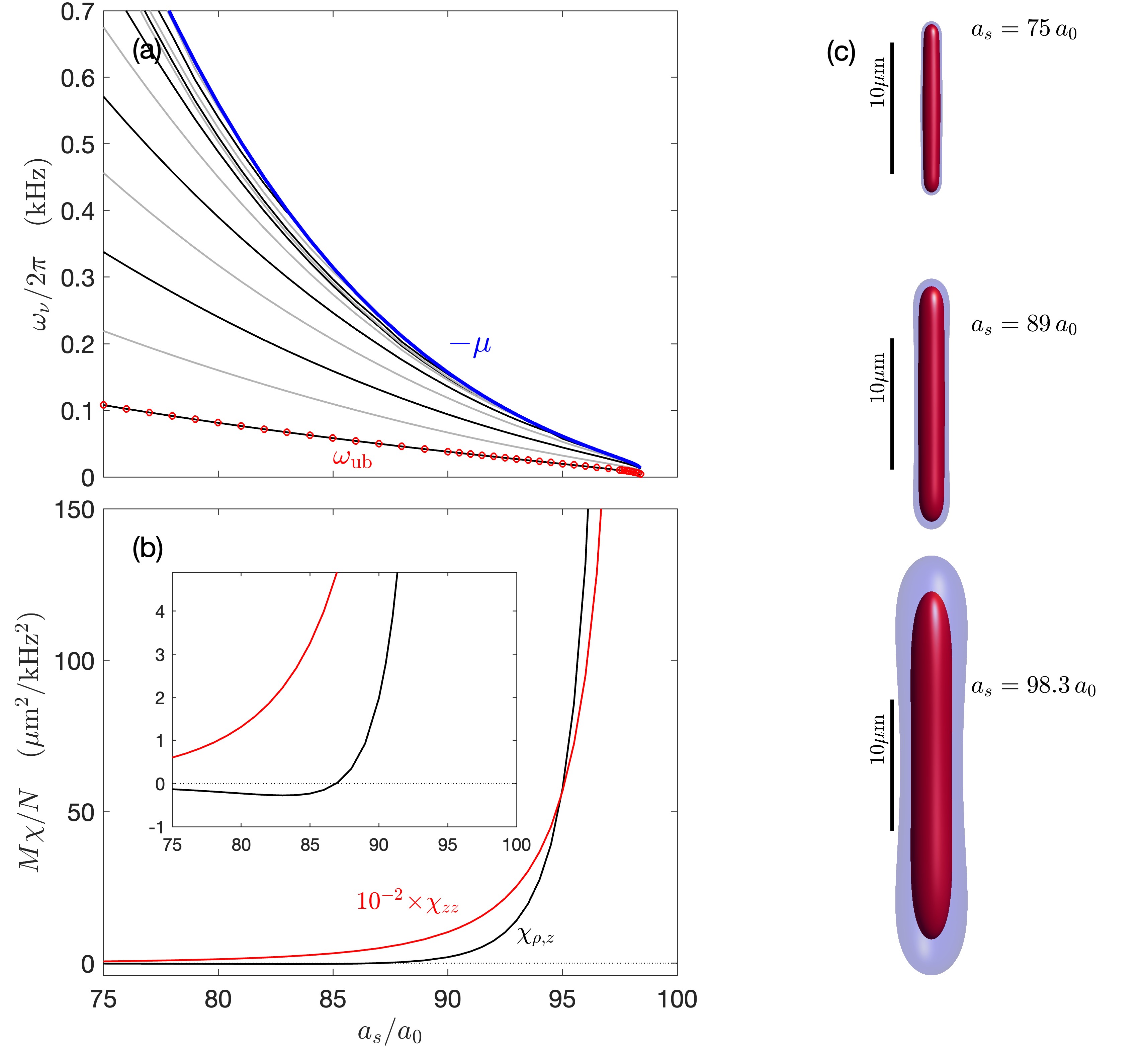} 
   \caption{Free-space droplet ($V=0$) as a function of s-wave scattering length $a_s$. (a) Results from numerical solution of the BdG equations for the $m=0$ excitations that are even (black lines) and odd (grey lines).  The droplet binding energy, characterized by $-\mu$, is shown for reference (blue line).
   The approximate result $ \omega_\text{ub}$ obtained from (\ref{omegaub}) (red circles). The droplet binding energy, characterized by $-\mu$, is shown for reference (blue line).
   (b) The static polarizabilities $\chi_{zz}$ and $\chi_{\rho z}$. 
(c)  Density isosurfaces of droplet states for various $a_s$ values as labelled. Isosurfaces shown for  $10^{18}\,$m$^{-3}$ (blue) and $10^{19}\,$m$^{-3}$ (red). 
Results for $N=10^4$ $^{164}$Dy atoms using $a_{\mathrm{dd}}=130.8\,a_0$.}
   \label{fig:fsdroplet2}
\end{figure*}

  In Fig.~\ref{fig:fsdroplet}(c) we show results for the static polarizabilities. The axial result $\chi_{zz}$ [used in Eq.~(\ref{omegaub})] is always positive, reflecting that the application of a weak axial confinement to the free-space droplet causes it to shorten. The magnitude of $\chi_{zz}$ tends to be much larger than $\chi_{\rho z}$, so we have scaled it by a factor of $10^{-2}$ to make the two polarizabilities easier to compare.  The transverse response $\chi_{\rho z}$ changes sign depending on the atom number. In the deeply bound droplet regime (i.e.~for $N>5\times 10^3$, where $\mu$ is large and negative), it is negative. Thus a weak compression along $z$ results in the droplet expanding transversally.  In contrast, closer to the unbinding threshold ($N<5\times 10^3$),  $\chi_{\rho z}$  is positive, and the transverse width will decrease with axial compression. This occurs as the lowest $m=0$ mode starts to soften.  This change in  behaviour of the transverse response occurs as the lowest excitation mode changes from quadrupolar character at high $N$ to monopole (compressional) character at low $N$ \cite{Baillie2017a}.

 \subsection{Results with varying $a_s$}
 
In Fig.~\ref{fig:fsdroplet2}(a) we show results for the $m=0$ excitations of a free-space droplet with a fixed number $N=10^4$, but with $a_s$ varying. Here the droplet unbinds at $a_s\approx98.5\,a_0$, when the lowest energy mode goes soft. The BdG calculations for the energy of the lowest mode are again seen to be in good agreement with the results of Eq.~(\ref{omegaub}). In  Fig.~\ref{fig:fsdroplet2}(b) we examine the response of the ground state to a change in axial confinement.  For the deeply bound droplet  ($a_s\lesssim  87\,a_0$) $\chi_{\rho z}$ is negative, while close to unbinding ($a_s\gtrsim  87\,a_0$) it becomes positive [see inset to Fig.~\ref{fig:fsdroplet2}(b)].

 \begin{figure*}[htbp!] 
   \centering
 \includegraphics[width=7in]{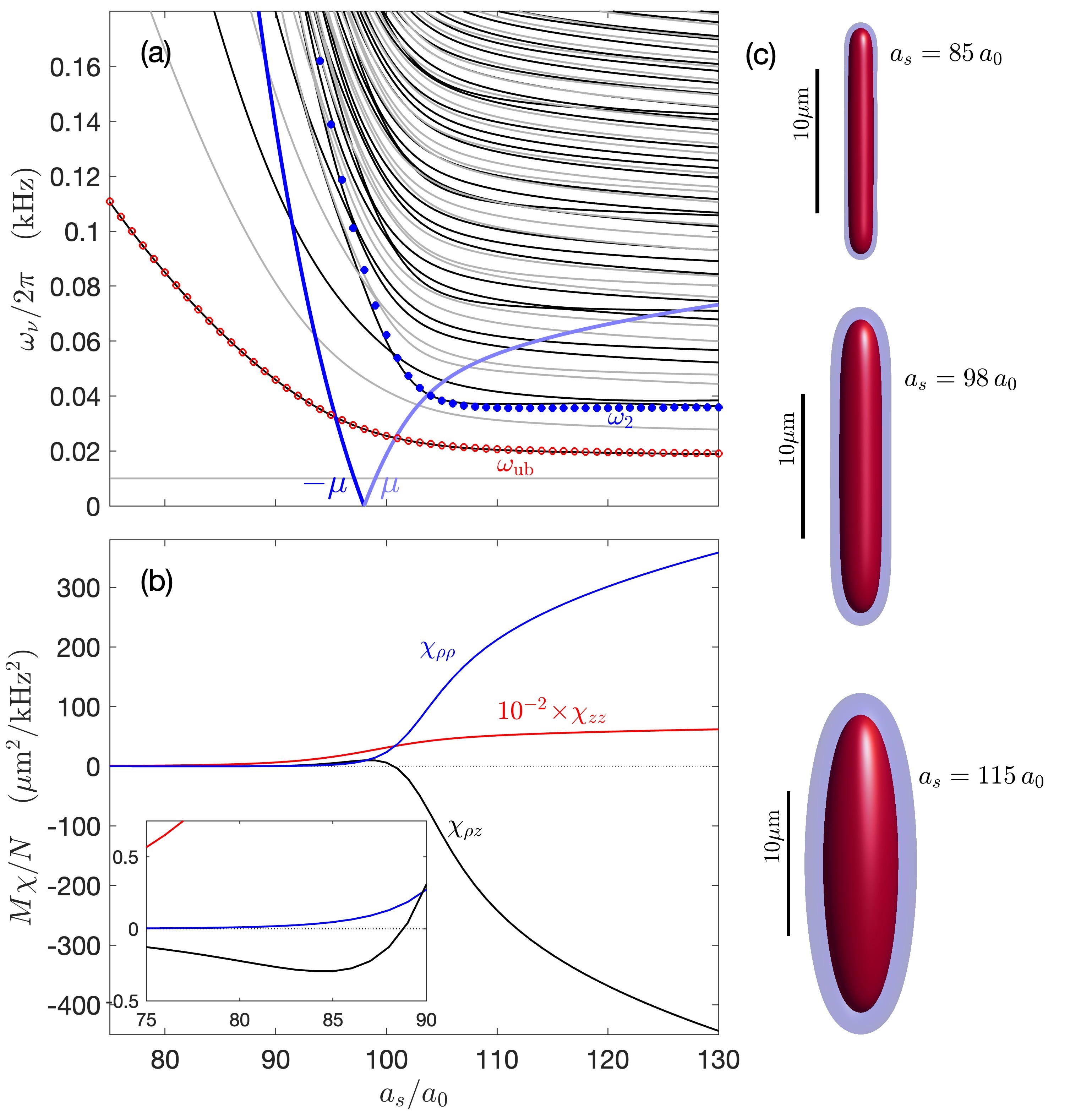}
   \caption{Trapped quantum droplet as a function of s-wave scattering length $a_s$. (a) Results from numerical solution of the BdG equations for the $m=0$ excitations that are even (black lines) and odd (grey lines).   The approximate result $ \omega_\text{ub}$ obtained from (\ref{omegaub}) (red circles) and the excitation frequency estimate $\omega_2$ from Eq.~(\ref{omega2}) (blue circles). The energy scales $\pm\mu$ are shown for reference (blue lines).     (b) The static polarizabilities $\chi_{zz}$, $\chi_{\rho z}$ and $\chi_{\rho\rho}$. 
(c) Density isosurfaces of droplet states for various $a_s$ values as labelled. Isosurfaces shown for  $10^{18}\,$m$^{-3}$ (blue) and $10^{19}\,$m$^{-3}$ (red). 
Results for $N=10^4$ $^{164}$Dy atoms using $a_{\mathrm{dd}}=130.8\,a_0$ and $(\omega_{x,y},\omega_z)/2\pi=(20,10)\,$Hz.} 
   \label{fig:trapdroplet}
\end{figure*}

 \section{Results for trapped droplet}
 In Fig.~\ref{fig:trapdroplet}(a) we show the excitation spectrum for a droplet confined in a weak cigar-shaped trap $(\omega_{x,y},\omega_z)=2\pi\times(20,10)\,$Hz. This geometry allows a smooth crossover between a self-bound droplet and a trap-bound dipolar BEC. Apart from the trap, the parameters of these results are otherwise identical to the case studied in Fig.~\ref{fig:fsdroplet2}. For $a_s\lesssim95\,a_0$ the lowest excitation energies, and chemical potential are similar for the trapped and untrapped cases. A notable feature is the emergence of a Kohn mode at the axial trap frequency\footnote{The transverse Kohn modes, with frequency $\omega_{x,y}$, are in the $m=\pm1$ excitation branches and do not appear here.} $\omega_z$. For $a_s\gtrsim95\,a_0$ the trap starts playing an increasingly important role. In this regime the droplet self-binding begins to fail and the trap provides the confinement  [see Fig.~\ref{fig:trapdroplet}(c)].  As this happens the chemical potential becomes positive. Since the trap binds excitations at all energy scales, we show BdG results with $\hbar\omega_\nu>-\mu$.

For the full range of parameters considered in  Fig.~\ref{fig:trapdroplet}(a)   Eq.~(\ref{omegaub}) is seen to provide an accurate description of the axial mode. The lower energy Kohn mode has no effect on our sum rule since the symmetry of the that mode does not couple tothe $\sigma_z$ operator. 
  In Fig.~\ref{fig:trapdroplet}(a) we also present results for the transverse equivalent of Eq.~(\ref{omegaub}), i.e.~
  \begin{align}
  \omega_2\equiv\sqrt{\frac{4N\langle \rho^2\rangle}{M\chi_{\rho\rho}}},\label{omega2}
  \end{align}
  where $\delta\langle\sigma_\rho\rangle=\lambda\chi_{\rho\rho}$ defines the static polarizability for a transverse perturbation of $-\lambda\sigma_\rho$. 
  The frequency estimate of Eq.~(\ref{omega2}) is in reasonable agreement with the second even mode in the trap-bound region, but ascends to a high frequency in the droplet regime. These results indicate that Eq.~(\ref{omega2}) is of limited use in the droplet regime. 
    
    We  show results for the static polarizabilities in Fig.~\ref{fig:trapdroplet}(b). In the droplet regime $\chi_{zz}\gg|\chi_{\rho\rho}|,|\chi_{\rho z}|$, while in the trap-bound regime they all become of comparable magnitudes. Interestingly $\chi_{\rho z}$ is positive at intermediate values of $a_s$.  Correspondingly we find that the axial mode changes character from being a quadrupolar excitation at low $a_s$ values, to monopolar at intermediate values, before returning to quadrupolar character at high values of $a_s$. Aspects of this behavior has also been described within a variational Gaussian approach \cite{Wachtler2016b}.

  \section{Conclusions}
  
 In this work we have presented results for the ground state properties of quantum droplets, their collective excitations and static polarizibility related to  a change in confinement. The lowest energy collective mode has been measured in experiments by  Chomaz \textit{et al.}~\cite{Chomaz2016a}, and plays a critical role in the instability of a free-space droplet. We have shown that a simple sum rule approach can accurately predict the frequency of this mode over a wide parameter regime. Our results for the static polarizabilities quantify changes in the widths of the droplet in response to variations in the axial or transverse confinement. We observe that the transverse static polarizability (to an axial confinement change) is negative in deeply-bound droplets and crosses over to being positive in weakly-bound or trap-bound droplets. This change correlates with lowest excitation changing from  quadrupolar (incompressible) to monopolar (compressible) character.

\section*{Acknowledgments}
PBB wishes to acknowledge use of the New Zealand eScience Infrastructure (NeSI) high performance computing facilities and support from the Marsden Fund of the Royal Society of New Zealand. Helpful discussions with D.~Baillie, A.-C.~Lee and J.~Smith are acknowledged.

\end{document}